# Smart-aggregation imaging for single molecule localization with SPAD cameras

Istvan Gyongy[1,*], Amy Davies[2], Neale A.W. Dutton[3], Rory Duncan[2], Colin Rickman[2], Robert K. Henderson[1] and Paul Dalgarno[2]

[1]*The University of Edinburgh, Institute for Integrated Micro and Nano Systems, Edinburgh, U.K.*
[2]*Heriot-Watt University, Institute of Biological Chemistry, Biophysics and Bioengineering, Edinburgh, U.K.*
[3]*STMicroelectronics Imaging Division, 33 Pinkhill, Edinburgh EH12 7BF, U.K.*
[*]*istvan.gyongy@ed.ac.uk*

**Abstract:** SPAD cameras offer single photon detection sensitivity, high frame rates and zero readout noise. They are a core technology for widefield FLIM, but have further potential in ultra-fast imaging applications. However, in practice sensitivity falls behind that of EMCCD and sCMOS devices due to the lower photon detection efficiency of SPAD arrays. This paper considers the application of a binary SPAD camera to the capture of blinking molecules as part of a superresolution (dSTORM/PALM) experiment. Simulation and experimental studies demonstrate that the effective sensitivity of the camera can be improved significantly by aggregation of signal only binary frames. The simulations also indicate that with future advances in SPAD camera technology, SPAD devices could in time outperform existing scientific cameras when capturing fast temporal dynamics.

## 1. Introduction

A technological trend in image sensors has been the push to achieve single photon resolution, through increased signal gain and electronic noise reduction [1]. Photo-cathode based intensifier imaging technologies, electron bombarding (EBCCD or EBCMOS) and intensifiers (ICCD and ICMOS) first showed wide-field low-light imaging for night vision and scientific applications [2]. Three solid-state imaging technologies have attained deep sub-electron read noise (DSERN) that combined with high enough optical sensitivity, have an inherent capability of single photon resolution: the electron multiplying CCD (EMCCD), CMOS Single Photon Avalanche Diode (SPAD), and two CMOS image sensors (CIS) devices namely the Pinned Avalanche Photo Diode (PAPD) [3], and the pinned photo-diode (PPD) [4].

The spatial resolution and optical fill factor of CMOS SPAD image sensors has progressively increased to be comparable to EMCCD and CIS through the reduction of in-pixel circuitry employing analogue counting and binary memory circuits [5]. The primary advantage of binary memory pixels over analogue solutions is the digital nature of operation, permitting fast data capture and transmission at >1G pixels per second and facilitating high frame rates at >10k frames per second (FPS) with negligible readout noise contributed from the digital electronics.

Single Molecule Localization Microscopy [6, 7] is one of the key applications in life science imaging requiring fast, high sensitivity cameras. The technique, which has many variants, uses either photo-activatable or stochastically activated fluorescent markers that are activated in spare subsets at a time. The resulting spatially distinct diffraction-limited point spread functions therefore be individually identified and localized, allowing high resolution images of localized emitters to be constructed over time. The technique has seen many developments (such as multi-color and 3D imaging [8]), but the challenge, from an imaging perspective, remains the same: to capture the millisecond molecule blinks with suitably high signal-to-noise ratio, so that their position may be estimated (typically by fitting a model of the point spread function) with the highest accuracy and precision.

In this context, a higher frame rate camera potentially allows the detection of the onset and duration of a blink to a greater degree, providing extra precision in time-windowing and signal intensity thresholding so that background noise may be suppressed and the integrated signal maximized. However, at sub-millisecond exposures there will be very few photons from the molecule, preventing molecules from being detected on single frames, let alone localized. Summing multiple sequential frames regains signal intensity, but the total readout noise increases with the square root of the number of summed frames, degrading the signal-to-noise ratio critical to super-resolution localization. This paper sets out a frame summing scheme, termed smart aggregation, that exploits the negligible read noise of SPAD cameras so that no additional noise is incurred. Averaging of frames in fixed periods degrades the temporal resolution, which motivates the use of a rolling time average to track molecule blinks. Once the blink onset and duration has been estimated, only frames deemed to contain the blink are summed for an optimized molecule image. Furthermore a model is presented to simulate optimal detector performance for localization microscopy, highlighting the key technological gains that, in addition to smart aggregation, are needed to further optimize the application of single photon array detectors.

The paper is organized as follows. Section 2 outlines the basic characteristics of a binary SPAD imager. In Section 3, the smart aggregation algorithm for composing molecule images is presented. Section 4 applies the algorithm in a super-resolution (STORM) experiment with nanorulers, and uses a simulation model to make projections as to the localization performance of future SPAD devices. Concluding remarks are given in Section 5.

## 2. Characteristics of a binary SPAD camera

The Quanta Image Sensor (QIS) concept [9] projects the recent image sensor developments of read noise reduction, decreasing pixel size (and diminishing full well) to an imaging array of single photon photodetectors with a binary response. The binary states of either 0 (no photon detected) or 1 (at least one photon detected) provide limited information, therefore these are summed in space and/or time to form a spatio-temporally oversampled greyscale image frame. A binary SPAD camera is an example of a QIS, its raw output consisting of spatial information of binary bits or "bit-planes", as illustrated in Fig.1., with each pixel producing a time-domain sequence of 1's and 0's.

A characteristic of a QIS is logarithmic compression in the response to light, akin to photographic film, due to the "pile-up" distortion of many incident photons recorded with the same logical high signal value as a single photon. This behavior has been predicted by theory [10], as well as verified experimentally using a SPAD imager [5]. Yet, in microscopy applications, the low light intensity, coupled by the high frame rate (short exposure per bit-plane) of SPAD devices, means that the number of photons detected per pixel per exposure is less than one (so there is little "pile-up distortion"), and as such one typically operates in the linear region of the response curve. Indeed, for low light applications, where one might have a few thousand incident photons/pixel/s, provided the frame rate is high enough (>1k FPS), binary pixel values, when aggregated together, are sufficient to reproduce the variations in light intensity in a scene.

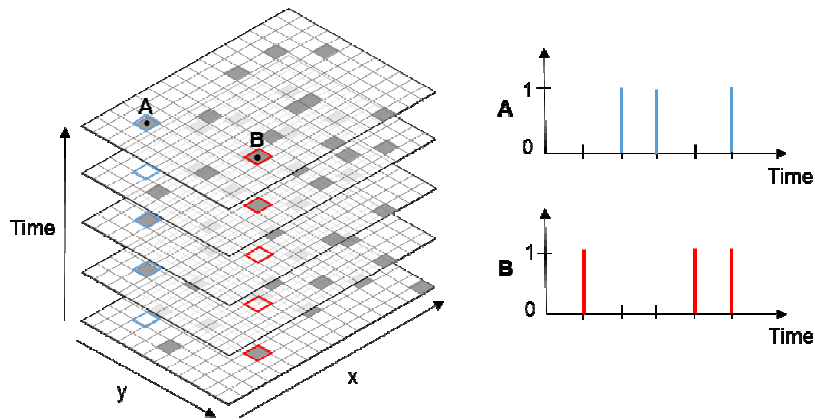

Fig. 1. Schematic representation if the output of a binary SPAD imager. Each pixel registers a 0 (no photon count) or 1 (photon count) in time.

The main noise source in a SPAD device is the pixel dark count rate (DCR), which refers to the spurious firings of the SPAD due to thermal events. The level of DCR depends on the pixel size and architecture, and will vary across the pixel array (as quantified using the parameter Dark Rate Non Uniformity, or DRNU). In a similar way, there will be a Photon Response Non-Uniformity (PRNU) in the photon detection efficiency (PDE) of individual pixels, that is the ratio between the average rate of photon detections and the rate of incident photons. Antolovic et. al. [11] discusses compensating for DCR, PRNU, and linearizing the response of SPAD QIS. Correcting for the DCR entails measuring the average count rate at each pixel, and subtracting the resulting "background" frame from subsequent images. The additional shot noise introduced by the dark count remains. Moreover, as in other camera technologies, SPAD imagers have a certain percentage of "hot pixels", where the DCR is much higher than average and masks true photon detections, rendering the pixels unusable.

Crucial to the operation of a SPAD imager in QIS mode is the fact that there is negligible read-out noise, which means that an arbitrary number of bit-planes may be summed without incurring a noise penalty. As discussed in the introduction, this represents a distinct advantage in the imaging of blinking molecules, in the context of Single Molecule Localization Microscopy. Fig. 2 illustrates the typically scenario when a molecule, switching on and off at random, is to be captured. The top graph shows the (simulated) light intensity trace from a single molecule, comprising a longer and a shorter blink. If imaging with a conventional EMCCD camera, integrating over fixed periods {A,B,C,D} (as indicated), then one would get the peak intensity values shown in the middle graph, and the frames shown on the right. The long blink would be captured mostly in frame B, with some of it temporally quantized into frame C, resulting in a faint image in that frame. Frame D would show an even fainter image of the short blink, due to the blink only appearing for a fraction of the integration period, and therefore being averaged out with the background. A binary SPAD camera captures bit-planes at a much faster rate than the frame rate of EMCCD. One could sum these bit-planes over fixed periods (as shown in the bottom graph), but the approach proposed here is to detect the 'on' times of the molecules, and then to sum bit-planes only over the 'on' periods. Noting that the lowest achievable localization uncertainty (see, for example, [12]) is largely dependent on the ratio of detected molecule photons over background photons, the suggested "smart aggregation" approach should yield optimized images from the perspective of localization.

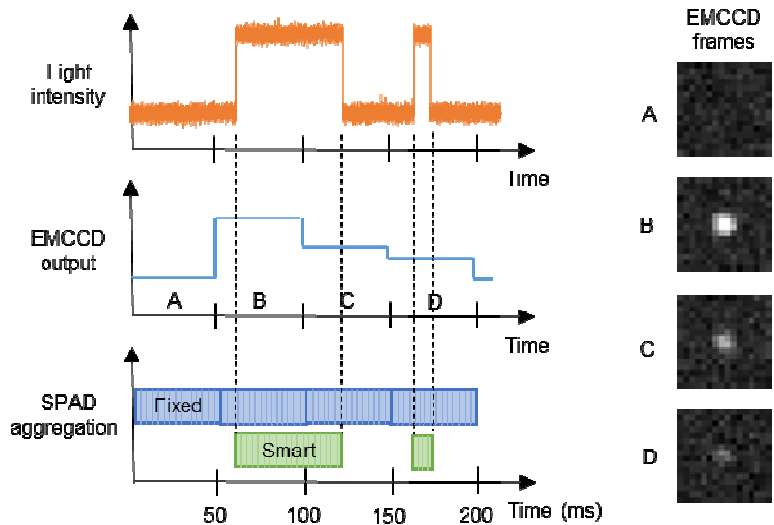

Fig. 2. Representation of the operational frame capture of an EMCCD and bit-plane imager for an example single blinking molecule. The top graph represents the molecule intensity trace in time, the middle graph the collected intensity from a fixed frame rate EMCCD (and shown pictorially in the right hand images), the bottom graph show the difference between fixed frame capture and smart-aggregated frame capture.

The SPAD sensor explored here (labelled SPCIMAGER) is a 320×240 resolution imager featuring a 8µm pixel pitch at 26.8% fill factor, and a peak photon detection probability of 35% at 450nm [13]. When operated in binary mode, bit-planes are captured at a rate of 10kfps. Both rolling and global shutter modes are available, but the former is typically used as it allows for back-to-back exposures at the maximum frame rate (so that the exposure time per bit-plane is 100µs). The imager is paired with an FPGA board (Opal Kelly XEM6310) that controls the acquisition of image data, relaying a continuous stream of bit-planes to a PC over a USB 3.0 link.

### 3. Smart aggregation

The steps in the bit-plane smart aggregation algorithm are outlined in Fig.3. The input to the system is the raw or unprocessed bit-plane images from the sensor. The first step is to apply Gaussian filtering in space to enhance any molecule flashes, and then time-averaging to track temporal behavior. The size (or width σ) of the Gaussian kernel is chosen so as to match the expected point spread function (PSF) of the molecule, which can be readily estimated from the optics of the microscopy and the wavelength in question. More specifically, $\sigma$ is chosen to be one-third of the Airy disk radius, which is estimated from $r=0.61\lambda/NA$ (not including the magnification), where $\lambda$ is wavelength and $NA$ is the numerical aperture of the objective. The spatial filtering is followed by time averaging, carried out with a rolling "window" [14], whose size corresponds to the shortest blink that can be reliably detected.

In the second step, thresholding is applied to the filtered frames to detect molecule flashes. Clusters of points are thereby identified as (potential) molecules, and in each case the local maximum of the filtered pixel values is used as a rough estimate $\{x_i,y_i\}$ of the molecule position, which then defines a region of interest (ROI) around the candidate molecule. Next, the thresholded time trace at $\{x_i,y_i\}$ is used to estimate the on and off time of the molecule flash in question.

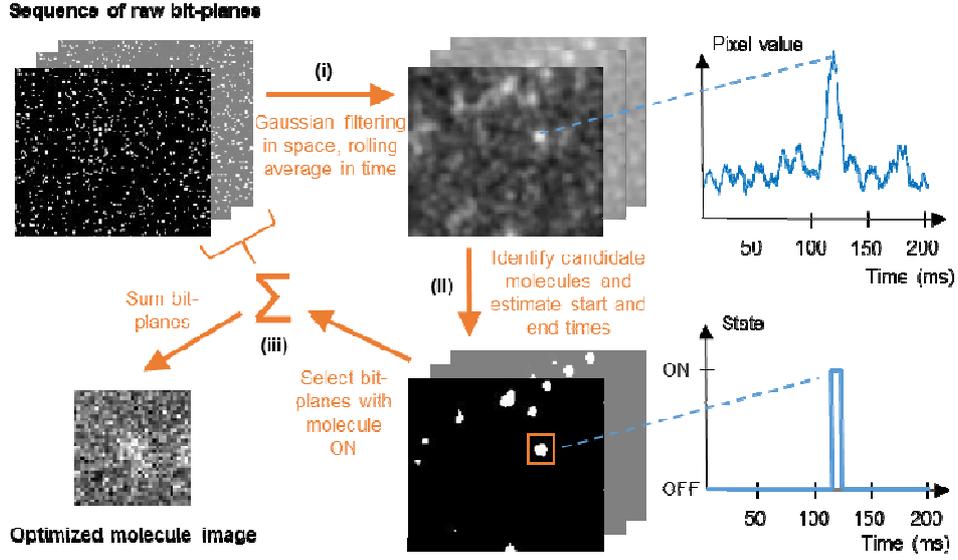

Fig. 3. The smart-aggregation scheme. A sequence of bit-plane images are (i) filtered, (ii) molecules are detected and positive photon events registered before (iii) summation of original bit planes according to aggregated photon arrival times.

The third step takes each candidate molecule in turn, selects the raw bit-planes deemed to contain the molecule flash (based on the estimated on and off times), and sums said bit-planes, cropped to the relevant ROI. As an end result one has a series of optimized molecule images, which can then be readily localized using standard methods [15]. The overall scheme is highly parallelizable and thus well suited to implementation on FPGA.

The duration of the time averaging window, and the threshold used for detecting molecules are derived from the underlying photon detection statistics and are a function of the photon flux from the molecule, the background photon level (including the dark count rate), as well as the effective pixel size (or optical magnification).

Consider a binary SPAD camera whose pixels have a uniform PDE $\eta$. For simplicity, each pixel is assumed to detect photons at the same average rate of $b$ times per second when subjected to background light only. It is further assumed that a molecule, when on, emits a photon flux of $I$ photons per second onto the sensor. The light from the molecule has a PSF centered around pixel $\{i,j\}$, and is approximated by a Gaussian with standard deviation $\sigma$ (normalized by the pixel size). It is also assumed that bit-planes are captured by the camera with an exposure time of $\tau$ and are filtered spatially with a $k \times k$ Gaussian filter (also of width $\sigma$), before being summed, in time, in groups of $N$. It can then be shown, using the Poisson statistics of photon arrivals, that the mean value of pixel $\{m,n\}$ (as seen on the raw bit-planes) is:

$$\mathrm{E}(B_{m,n}) = \begin{cases} 1 - \exp(-b\tau) & \text{when molecule is OFF} \\ 1 - \exp(-IG_{m,n}\eta\tau - b\tau) & \text{when molecule is ON} \end{cases}, \quad (1)$$

where $G$ is the Gaussian PSF, as sampled (discretized) over the pixel array. Similarly, the variance of the pixels can be expressed as:

$$\text{Var}(B_{m,n}) = \begin{cases} \exp(-b\tau)(1-\exp(-b\tau)) & \text{when molecule is OFF} \\ \exp(-IG_{m,n}\eta\tau - b\tau)(1-\exp(-IG_{m,n}\eta\tau - b\tau)) & \text{when molecule is ON} \end{cases}. \quad (2)$$

Now provided the pixel values $B$ can be treated as independent random variables, which are uncorrelated in time (within an on or an off period), the pixel $\{i,j\}$ in the filtered $F$, and aggregated image $H$ will have the following mean and variance:

$$\text{E}(H_{i,j}) = N\,\text{E}(F_{i,j}) = N \sum_{m=i-(k-1)/2}^{i+(k-1)/2} \sum_{n=j-(k-1)/2}^{j+(k-1)/2} G_{m,n}\,\text{E}(B_{m,n}), \quad (3)$$

$$\text{Var}(H_{i,j}) = N\,\text{Var}(F_{i,j}) = N \sum_{m=i-(k-1)/2}^{i+(k-1)/2} \sum_{n=j-(k-1)/2}^{j+(k-1)/2} G_{m,n}^2\,\text{Var}(B_{m,n}). \quad (4)$$

As $N$ increases, and one is summing more and more pixel values $B$ to compose $H$, the distribution of $H$ will tend to a Gaussian distribution, according to the Central Limit Theorem. The mean and variance of this distribution will depend on whether the molecule is on or off. Thus to ensure that the molecule can be reliably detected, the probability distributions in the on and off cases have to be suitably separated. If one is to aim for a ~99% detection accuracy based on the value of $H$ at $\{i,j\}$ (assuming the molecule was either on or off over the whole of the aggregation period), then the point where the tails of the two distributions intersect should be three standard deviations away from the means of the distributions, so the requirement is:

$$\text{E}(H_{i,j}^{OFF}) + 3\sqrt{\text{Var}(H_{i,j}^{OFF})} = \text{E}(H_{i,j}^{ON}) - 3\sqrt{\text{Var}(H_{i,j}^{ON})}. \quad (5)$$

Thus, combining Eq. (5) with Eq. (3) and (4) one can show that the minimum number of bit-planes to be summed is:

$$N = 9\frac{\left(\sqrt{\text{Var}(F_{i,j}^{OFF})} + \sqrt{\text{Var}(F_{i,j}^{ON})}\right)^2}{\left(\text{E}(F_{i,j}^{ON}) - \text{E}(F_{i,j}^{OFF})\right)^2}, \quad (6)$$

and the corresponding detection threshold $T$ can be expressed as:

$$T = N\,\text{E}(F_{i,j}^{OFF}) + 3\sqrt{N\,\text{Var}(F_{i,j}^{OFF})}. \quad (7)$$

Now in practice $H$ is a rolling sum of the filtered bit-planes $F$, so that at time step $t$,

$$H_{i,j,t} = F_{i,j,t} + F_{i,j,\{t-1\}} + \ldots + F_{i,j,\{t-N+1\}}. \quad (8)$$

The precision with which one can determine the start and end times of the molecule flash will be dependent on the length of summation. One approach is to take the time corresponding to the middle of the sequence of $N$ bit-planes when $T$ is crossed (leading to an uncertainty in the estimated start and end times $\{t_s, t_e\}$ of approximately $\pm N/2$ bit-planes). When composing the optimized molecule image, the sum of bit-planes is instead carried out from $t_s - N/2$ to $t_e + N/2$ to reduce the chances of bit-planes with useful signal being discarded. To avoid a single molecule flash being treated as multiple flashes, instances of $H_{i,j}$ dropping below $T$ for a single frame only are ignored when establishing $\{t_s, t_e\}$.

A corollary of the above analysis is that the higher the anticipated photon flux from the molecule (or the PDE $\eta$ the camera), the lower $N$ can be, and hence the better one can establish the time history of the molecule flash (as well as being able to detect shorter on or off periods).

**Table 1. Details of the camera specification as used in the model.**

| Type | Binary SPAD | EMCCD | sCMOS |
|---|---|---|---|
| Model | SPCIMAGER (& projected) | Andor iXon Ultra 987 | Hamamatsu ORCA-Flash4.0 V2 |
| Pixel size | 8 μm | 16 μm | 6.5 μm |
| Quantum Efficiency | 35% PDE | 90% | 80% |
| Fill factor | 26% (78%) | N/A | N/A |
| Read noise | Negligible | 0.2 e$^-$ (input referred) | 1.4 e$^-$ |
| Dark Noise | 100Hz (25Hz) median DCR | 0.001Hz DCR | 0.05Hz mean DCR |
| Other Noise | N/A | 0.0018Hz CIC | N/A |
| Non-uniformity | 2% DRNU, 1% PRNU | N/A | 1% DRNU, 0.5% PRNU |
| Frame time | 100μs per bit-plane | 50ms | 50ms |

## 4. Application of smart aggregation

To assess the benefit of smart aggregation experimental studies were carried out in which the localization uncertainty resulting from the algorithm, as applied to SPCIMAGER, was compared with fixed aggregation and a reference EMCCD camera. This was followed by further simulations involving a projected, future SPAD camera, which was compared with current EMCCD and sCMOS devices to assess the future potential of smart aggregation bit-plane technology. An overview of camera characteristics, as assumed in the simulations, is given in Table 1.

The statistical models used here to simulate the image capture process of the cameras are similar to those in the comparison study of [16]. In particular, photon detections and dark noise (in all three cameras) are assumed to be Poisson processes, the electron multiplication stage of the EMCCD is modelled as a Gamma process, and read noise is Gaussian distributed. An important difference in the simulations here is that molecule undergoes stochastic blinking so that the principle of smart aggregation may be tested.

For the simulated SPAD camera a 32×32 pixel region of interest (ROI) is assumed; for the EMCCD and sCMOS camera models the ROI is 16×16 and 40×40, respectively (to ensure similar fields of view). The simulated molecules switch on, one at a time, at random times and locations, and with a given blink duration. As in the analysis of Section 3, the molecule PSF is approximated by a Gaussian function.

*4.1 Experimental results*

In the experiments, a sample of GATTA-PAINT 80G nanorulers [17] was captured using SPCIMAGER and a Hamamatsu ImageEM EMCCD camera. The sample consist of triplets of fluorescent markers, based on the ATTO 550 dye, that are 80nm apart, and is intended to be used as a calibration slide for superesolution microscopy systems. Both cameras were coupled to a Olympus Cell Excellence IX81 microscope operated in a STORM configuration, with a 561 nm excitation laser in TIRF illumination and a 150× TIRF objective. A 50/50 non-polarizing beam splitter unit (TwinCam by Cairn Research) was used to direct the image onto each camera for simultaneous imaging.

In the first test, SPCIMAGER and the EMCCD device were used sequentially to image different fields of the GATTA-PAINT nanorulers. In the case of SPCIMAGER, frames were generated by two different means: by summing non-overlapping groups of 640 bit-planes (to match, approximately, the EMCCD's frame time of 64ms) and by using the smart aggregation scheme described in Section 2. The background and molecule intensity levels inferred from the camera outputs were fed into the simulation models and used to generate corresponding sets of simulated images. Both the experimental and simulated images frames were then localized using Maximum Likelihood fitting, via the widely-used ThunderSTORM ImageJ plugin [18]. The localization errors obtained from the simulated data (for which the ground truth molecule positions are known) were compared with the localization uncertainties reported by ThunderSTORM for the true EMCCD and SPCIMAGER frames, as a means of validating the simulations models.

Fig 4. shows the results, the left graph plotting the simulated and experimental data for SPCIMAGER (fixed and smart aggregation) and the right graph showing the EMCCD data sets are presented. The vertical axes represent the root mean square (RMS) error in the localization, or, in the case of the experimental data, the combined uncertainty (also a RMS quantity). The uncertainty and error values are plotted against a range of blink durations, as prescribed in the simulations, and determined approximately from the experiment data (based on SPCIMAGER's aggregation algorithm or the number of consecutive EMCCD frames that the molecule is present, such that the localizations are within 40nm from one another on subsequent frames). The measured blink durations, in the case of SPCIMAGER, are rounded to the nearest 10ms so that experimental data points with similar blink durations may be combined and overall uncertainty figures calculated. Similarly, for the EMCCD data, combined uncertainty values are computed, with data points being grouped according to the number of frames that the molecule is emitting. In addition, a separate set of uncertainty values are obtained, based on merging localizations deemed to result from the same molecule blink (to produce a comparable data set to SPCIMAGER smart aggregation). It should also be noted that the EMCCD camera does not expose throughout the frame time; there is a certain read out time, in this case 30ms, during which light is not collected so a proportion of the shorter blinks will invariably be missed.

The results in Figure 5 show that for the SPCIMAGER smart aggregation of the bit-planes offers a clear improvement in localization compared with fixed aggregation, with the uncertainty reducing by for a factor of two for longer blink durations. For the EMCCD merging localizations also result in a reduction in uncertainty, though to a much lesser extent of around 20%. At short blink durations the improvement is reduced, largely due to the reduced photon numbers globally limiting localization certainty. The simulated data matches well to the experimental data, providing confidence in the simulation. The small discrepancy between the simulations and experiment is largely due to the variability in the blink intensity being unmodelled in the simulations. Furthermore, ThunderSTORM's uncertainty estimates (for the experimental data) do not take into account the non-uniformity in photon response and dark count across the SPAD array (and in the case of EMCCD, the software does not account for read noise).

It is interesting to note in the data presented in Figure 5, that SPCIMAGER outperforms the EMCCD camera for long blink durations. The reason for this is that long blinks, spreading over multiple EMCCD frames, result in several localizations, some of which can be imprecise due to reduced photon counts (when the molecule blink just spills into an extra frame) and may not be recognized as coming from the same molecule. The additional "poor" localizations lower the overall precision, though they can be avoided, to some extent, by increasing the detection threshold (or filtering localizations based on the estimated uncertainty). This problem does not arise for smart-aggregated bit-plane imaging. Fig 5. compares histograms of the blink duration, as estimated from the experimental data from the two cameras (via the smart aggregation scheme, and counting the number of consecutive EMCCD frames showing the same apparent blink). The two histograms are similar in shape, though SPCIMAGER provides higher time resolution, allowing the decay in the apparent distribution of blink durations to be observed in more detail.

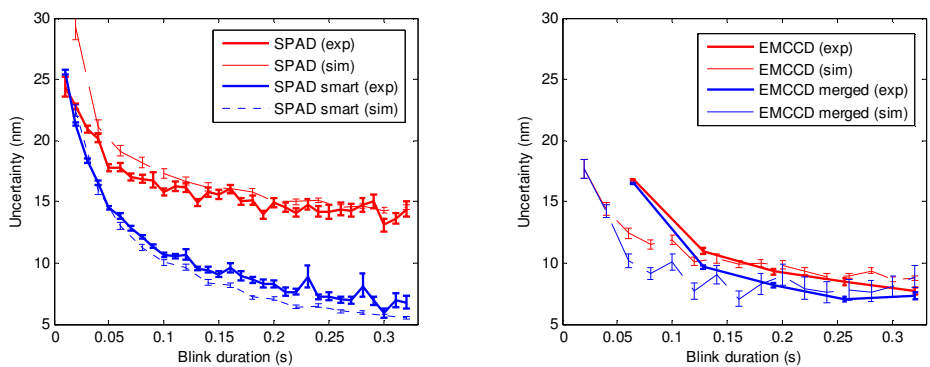

Fig. 4. Single molecule localization uncertainty for varying molecule blink durations as determined for SPCIMAGER with and without smart-aggregation and the EMCCD with and without merged frames. Shown are both experimental data from GATTA-Paint 80G rulers and simulated data.

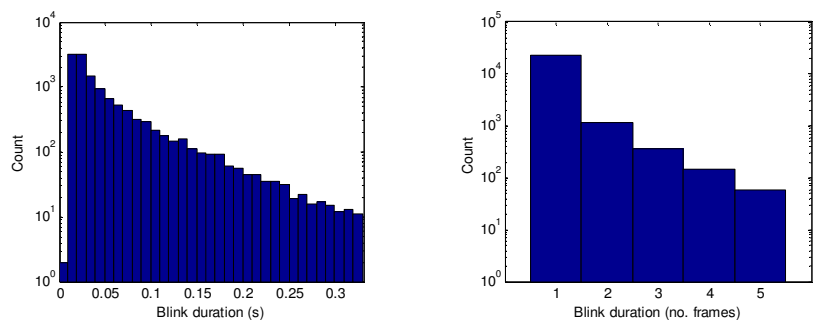

Fig. 5. Histograms of estimated experimental blink duration of single molecules, as obtained from SPCIMAGER (left) and EMCCD (right).

To verify the localization accuracy and precision, SPCIMAGER and the EMCCD device captured the same field of view of a GATTA-PAINT nanoruler sample *simultaneously* for 20 minutes using the TwinCam 50/50 unit. Fig. 6 shows an example of one nanoruler, as localized from EMCCD, SPCIMAGER fixed frame, and SPCIMAGER smart frames. The localizations are visualized using the normalized Gaussian option of ThunderSTORM (left plots), whereby each localization contributes to the rendered image in the form of a Gaussian function with standard deviation equal to the estimated uncertainty. ThunderSTORM's density filter has been applied to filter out localizations with fewer than three neighbors within a 30nm radius, so as to remove "lone" localizations resulting from noise or autofluorescense. Note that as each constituent marker within the nanoruler blinked multiple times during the imaging period, every marker is reproduced as a scatter of localizations, with the scatter relating to the density of localizations. The right plots show the cross-section of the Gaussian-rendered localization map of the triplet Gatta system. Peak-to-peak distances map well with the calibrated 80 nm GATTA ruler sample.

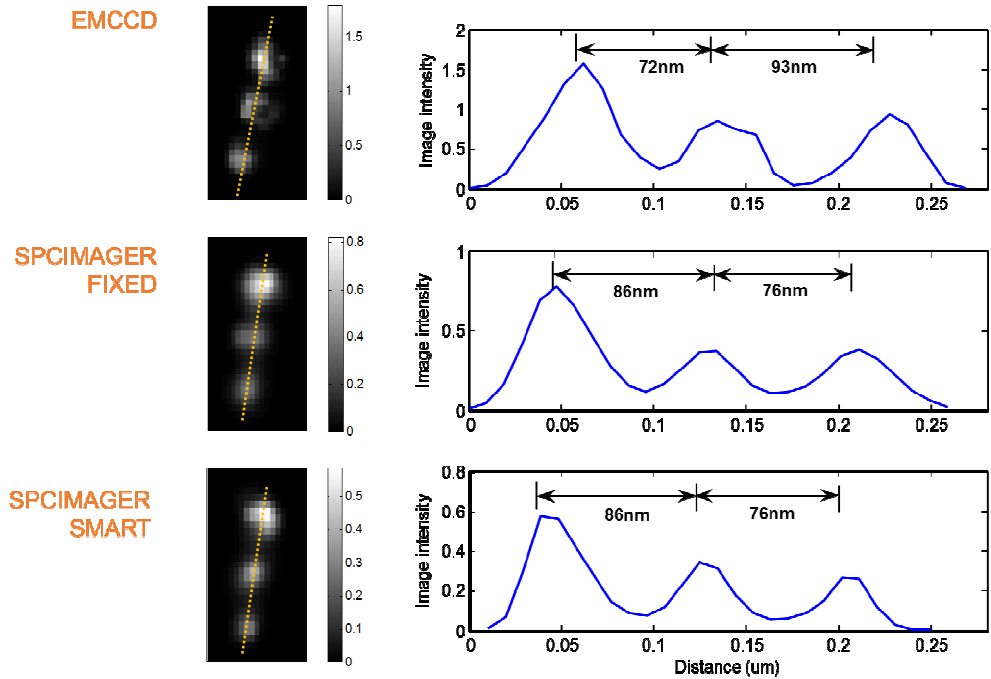

Fig. 6. Single Molecule STORM localizations of a single GATTA-PAINT 80G nanoruler as captured by EMCCD and SPCIMAGER, analyzed as both standard and smart aggregation.

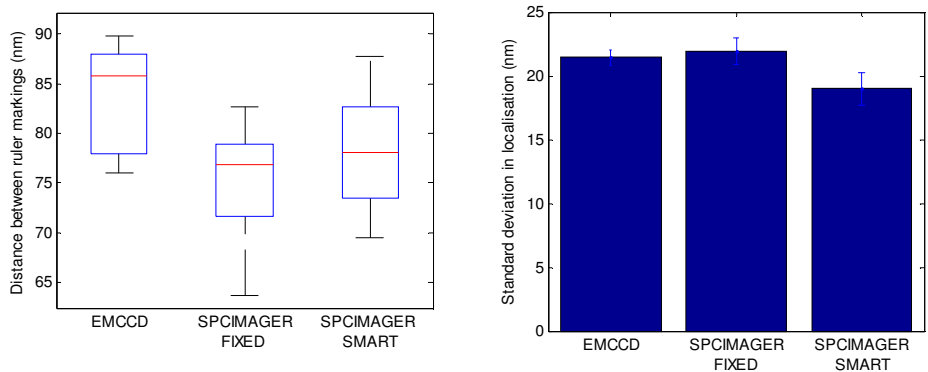

Fig. 7. Statistics of nanoruler localization: accuracy and standard deviation

The standard deviation of the localization is seen to reduce as a result of smart aggregation, leading to a "clearer" separation between the nanoruler markers. A summary of the localization statistics for this nanoruler and four others, calculated for all three imaging schemes, is shown in in Fig. 7. On the right are box plots of the observed distances between neighboring markers as measured using the mean localizations of each of the markers; the left graph gives the overall standard deviation in the localization of the markers. The presented results suggest that the localizations derived from the EMCCD and SPCIMAGER frames have comparable levels of. On the other hand markers localized from the SPCIMAGER's smart aggregated frames have greater precision due to the optimized nature of the frames leading to more consistent localizations.

*4.2 Projections for future SPAD device*

To assess the future capabilities of a SPAD imager, it was assumed that the effective fill factor will, in time, increase three-fold, with the DCR reducing by a factor of four. It is believed that these are realistic assumptions; there are a number avenues in which such a fill factor improvement may be realized, including microlensing [19], back-side illumination [20] or stacking [21]. Moreover, experimental studies suggest that SPADs with a similar structure to the sensor studied here exhibit a halving of DCR for every 8°C of cooling until around -10°C [22]. Thus it is anticipated that a four times reduction in DCR is achievable with a moderate level of cooling.

As before, the simulations considered a range of molecule blink durations, and for each duration 500 images were generated, localized, and the individual localization errors (distances to ground truth) combined to form an overall mean square error. Localisations further than 53.3nm from the ground truth (corresponding to the effective pixel size with a 150× objective) were considered to be false detections. The photon flux from the molecule was taken to be 50000 photons/sec, with a background level of 70000 photons/sec for every $\mu m^2$ area in the sample (under the assumption of TIRF conditions). Furthermore, the microscope was assumed to feature a 150× (in the default case) or a 60× objective.

The results are given in Fig. 8. The top-left graph compares the localization error resulting from the current SPAD (in other words, SPCIMAGER), with the projected, future SPAD. Both fixed and smart frame aggregation are considered, and dotes lines are used to represent regions where the molecule detection rate is below 90%. The error is seen to (approximately) halve with the future SPAD device. In the top-right graph, the future SPAD (with smart aggregation) is compared with EMCCD and sCMOS. As previously, we consider the effect of merging consecutive EMCCD/sCMOS localizations if within 40nm of each other. The results show the future SPAD camera matching the reference for most blink durations, apart from around 20-30ms, in which case sCMOS gives better results. The bottom-right graph shows the effects of switching to the 60× objective, in the cases of the current and future SPAD camera. Smart aggregation is assumed throughout. It is noted that the 60× objective reduces the localization error of the current SPAD by a moderate extent (by around 20% for longer blink durations), but has negligible effect on the future SPAD. The reason is that with the future camera, the DCR is no longer significant; the error is mainly caused by the background photon count, which simply gets re-distributed across the pixel array as the magnification is changed (in a similar way, the simulated error curve for the sCMOS was found to be largely unaffected by the change in assumed objective).

The bottom right graph in Fig. 8. considers the effect of an increasing number of hot pixels in a SPAD camera. Hot pixels have thus far not been modelled here, but their percentage can be considerable in an uncooled SPAD (in the region of 10% for longer exposure times). For this simulation, the model for the current SPAD was used (50ms total exposure was assumed with the molecule being on throughout). The simulated images were localized using the Maximum Likelihood estimator code of [23], which was modified to allow for undefined pixel values. The graph plots the results of two different strategies for handling hot pixels: interpolating over them (using the method of [24]) and ignoring hot pixels altogether in the fitting. Both 60× and 150× objectives are considered. In both cases, ignoring hot pixels appears to be the preferable option, and fewer than around 10% hot pixels does not seem to affect the localization error significantly (most of the error originating from the DCR and background photons).

## 5. Conclusion

Binary SPAD cameras, operating at a high frame rate, offer flexibility in the imaging of blinking molecules. A technique, termed "smart aggregation" has been presented here, which, for every detected molecule, sums only those binary fields when the molecule is on, for optimized images. The advantage of the approach has been demonstrated in both simulations and experiments involving nanorulers, with localization errors reducing considerably. The simulations suggest that with the anticipated improvement in fill factor and reduction in dark noise, a future SPAD imager featuring smart aggregation will match or outperform existing sCMOS and EMCCD cameras in molecule localization.

Whilst the present work uses standard localization algorithms, a camera-specific scheme (as in [25] for sCMOS cameras), which takes into account the non-uniformity in the pixel array, would likely perform better. It is expected that such as scheme, when applied to SPAD data, would result in higher accuracy, and a more symmetric distribution of localizations around the ground truth.

**Acknowledgements**

This research was funded by the ERC TotalPhoton grant. The authors appreciate the support of STMicroelectronics who fabricated SPCIMAGER. The use of the ESRIC facilities at Heriot-Watt University is also gratefully acknowledged.

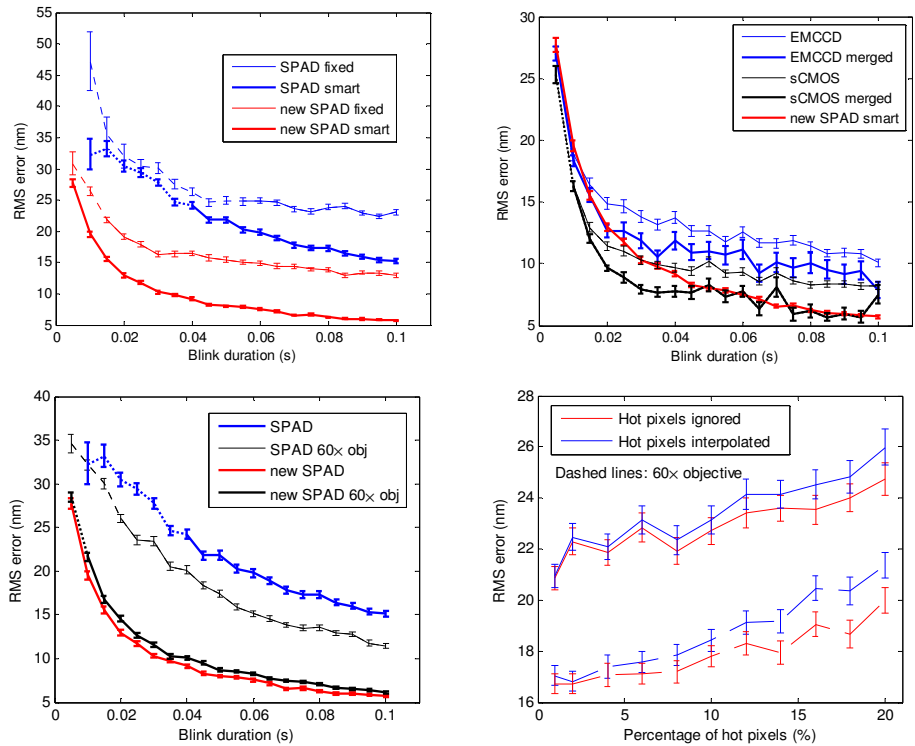

Fig. 8. Projections from simulation model: localization performance of current SPAD versus future SPAD (top-left), future SPAD as compared with sCMOS and EMCCD (top-right), the effect of 60× objective on the localisation of current and future SPADs (bottom-left), and the effect of hot pixels on the localisation error of current SPAD (bottom-right).